\documentclass[12pt]{article}
\usepackage{overcite}
\usepackage{graphicx}
\begin{document}

\title{Effect of ion hydration on the first-order transition in the
sequential wetting of hexane on brine}
\begin{small}
\author{Volker C.\ Weiss\thanks{E-mail: volker.weiss@fys.kuleuven.ac.be}
        \, and Joseph O.\ Indekeu\\[0.5cm]
        Laboratorium voor Vaste-Stof\/fysica en Magnetisme,\\
        Katholieke Universiteit Leuven, B-3001 Leuven, Belgium\\[0.3cm]}
\date{(February 27, 2003)}
\maketitle
\end{small}

\begin{abstract}
\noindent In recent experiments, a sequence of changes in the wetting
state (`wetting transitions') has been observed upon increasing the
temperature in systems consisting of pentane on pure water and of hexane on 
brine. 
In this sequential-wetting scenario, there occurs a first-order transition
from a partial-wetting state, in which only a microscopically thin film of
adsorbate is present on the substrate, to a `frustrated complete wetting state'
characterized by a mesoscopically, but not yet macroscopically thick wetting 
film. At higher temperatures, one observes a
continuous divergence of the film thickness and finally, at the 
critical-wetting temperature, the complete-wetting state, featuring a
macroscopic film thickness, is reached. 
This sequence of two transitions is brought about by an interplay of 
short-range and long-range interactions between substrate and adsorbate.
The critical wetting transition is controlled by the long-range forces
and is, thus, found by determining where the Hamaker constant, as
calculated from a Dzyaloshinskii--Lifshitz--Pitaevskii-type theory,
changes sign. The first-order transition involves both short-range
and long-range forces and is, therefore, more difficult to locate. 
While the pentane/water system is well understood in this respect by now, 
a detailed theoretical description of the hexane/brine system is hampered 
by the {\em a priori}\/ unknown modification of the interactions between 
substrate and adsorbate upon the addition of salt. 
In this work, we argue that the short-range interaction (contact energy)
between hexane and pure water remains unchanged due to the formation of 
a depletion layer (a thin `layer' of pure water which is completely devoid
of ions) at the surface of the electrolyte and that the presence
of the salt manifests itself only in a modification of the long-range 
interaction between substrate and adsorbate. 
In a five-layer calculation considering brine, water, the first layer
of adsorbed hexane molecules, liquid
hexane, and vapor, we determine the new long-range interaction of brine
with the adsorbate {\em across}\/ the water `layer'. According to
the recent theory of the excess surface tension of an electrolyte by
Levin and Flores-Mena, this water `layer' is of constant, i.e.\ 
salt-concentration independent, thickness $\delta$, with $\delta$ being
the hydrodynamic radius of the ions
in water. Once this radius has been determined, the first-order transition
temperatures can be calculated from the dielectric properties of the 
five media. Our results for these temperatures are in good agreement
with the experimental ones. 
\end{abstract}
\noindent PACS: 05.70.Fh; 68.10.-m; 68.10.Cr; 68.45.Gd \\
\noindent Keywords: Wetting, Surface tension, Electrolytes
\vspace{1cm}

\section{INTRODUCTION}
Imagine a volatile substance, the adsorbate, to be at liquid--vapor coexistence 
and in contact with a third phase, the substrate, which might be solid or
liquid. There are two qualitatively different ways in which the three
phases can meet: one possibility, known as partial wetting, is that the
liquid phase of the adsorbate forms discrete droplets on the surface of
the substrate, which have non-zero contact angle with this surface. 
In this case, the 
droplets are connected only by a microscopically thin film of adsorbate 
that covers the substrate. According to Young's equation \cite{rowl}, the 
contact 
angle $\theta$ is related to the three interfacial tensions between
substrate and liquid ($\gamma_{sl}$), substrate and vapor ($\gamma_{sv}$), 
and liquid and vapor ($\gamma_{lv}$) and can be calculated from $\gamma_{sv}
- \gamma_{sl} = \gamma_{lv} \cos{\theta}$. 
The second possible arrangement of the three phases is that the liquid
phase forms a macroscopically thick layer on the substrate surface
in such a way that there is no direct contact of substrate and vapor
anymore. This situation is known as complete wetting and the three
interfacial tensions obey Antonow's rule: $\gamma_{sv} - \gamma_{sl} = 
\gamma_{lv}$. From this relation, it follows immediately that the
contact angle $\theta$ is zero for complete wetting. In his seminal
article on this subject, Cahn \cite{cahn} demonstrated the possibility of a 
transition
between the two states, partial and complete wetting, as, for example, the
temperature is varied. He predicted that, close enough to the liquid--vapor
critical point of the adsorbate, there will be a wetting transition from
partial to complete wetting (or drying). His theory also predicts this 
transition to be of
first-order nature -- by now, it has been shown experimentally that this
is usually the case for wetting transitions
\cite{moldover,wegdam,rutledge,cheng,kellay}.  
The discontinuous nature of the transition manifests itself in the abrupt
jump of the layer thickness at the transition temperature $T_{w,1}$, in the
existence of a prewetting line \cite{rutledge,cheng,kellay}, and
in the occurrence of hysteresis, one of the hallmarks of first-order 
transitions \cite{wegdam,rutledge}.

Nevertheless, since the early days of the theory of wetting, there had been
speculations about the possibility of a critical
(or higher-order) wetting transition, in which the wetting-layer thickness
diverges continuously to a macroscopic value
\cite{nakanishi,kroll,dietrich,brezin,lipowsky,nightingale,fisher}.
In this context, it is important to distinguish between long-range and 
short-range critical wetting. For the latter, the film thickness is
predicted to diverge logarithmically \cite{nakanishi,brezin,fisher}
(within mean-field theory), whereas, for the
former, a power-law behavior according to the relation $l \sim \Delta T^{-1}$, 
where $l$ is the film thickness and $\Delta T = T_{w,c} - T$ measures the 
distance from the critical-wetting temperature, is expected
\cite{kroll,dietrich,nightingale,shenoy}. 
While the upper critical dimensionality for long-range critical wetting
is less than three, which implies that
the predictions of a mean-field theory are valid, it is equal to three
for short-range critical wetting \cite{brezin,lipowsky,nightingale}. In 
contrast to long-range critical wetting, the occurrence of short-range critical
wetting requires the proximity of a bulk critical point \cite{nakanishi,brezin}. 

While short-range critical wetting has been seen very recently in 
methanol/$n$-nonane mixtures \cite{ross}, the first experimental 
observation of critical wetting was 
for an example of the long-range type: Ragil {\em et al.}\/ reported a 
continuous divergence of the film thickness for pentane on water 
\cite{ragil}. From the effective exponent of $-1$ that describes this 
divergence (see above) and the fact that the location of the transition
(53$^{\circ}$C) coincides with the temperature at which the Hamaker constant
changes sign, it was concluded
that long-range forces bring about the critical transition \cite{ragil}.
In accord with this assumption is the considerable distance of the critical
wetting temperature from any bulk critical point.

For the system of hexane on pure water, i.e.\ for a three-layer configuration 
of the form water/liquid hexane/vapor, the critical-wetting temperature was 
estimated -- based on calculations of the Hamaker constant -- to be 
$T_{w,c} = 96^{\circ}$C and is, therefore, too high to be observed using
the existing experimental set-up \cite{shahidzadeh}. In order to depress the 
(critical-)wetting temperature, salt (NaCl) was added to the system.
\cite{shahidzadeh,richmond,dussaud}
Interestingly, for salt concentrations of 1.5 mol/L and 2.5 mol/L, {\em
two}\/ marked changes of the film thickness were observed ellipsometrically
\cite{shahidzadeh}:
at low temperatures, the adsorbed film is only a few {\AA} thick; on
passing the temperature $T_{w,1}$, there is an abrupt increase of the film 
thickness
to about 100 {\AA}. The hysteresis which is observed for this transition
corroborates that this change of the wetting state is of first-order nature.
Upon increasing the temperature further, the film thickness grows
continuously and diverges at $T_{w,c}$, just as it had been observed for
pentane on water. Returning to the pentane/water system, it was found that,
when heating the system from low temperatures instead of cooling it down
from $T_{w,c}$, as had been done before, a first-order thin--thick transition 
occurs at 25$^{\circ}$C in this system as well \cite{bertrand}.

On the basis of the conventional Cahn theory, the first-order transition in
this system was 
predicted to occur at $-30^{\circ}$C \cite{ragil2}; a modification of the
original theory by Dobbs \cite{dobbs}, however, which treats the first 
layer of adsorbate
molecules in a lattice-gas approximation and, thereby, fulfills Henry's law, 
brought
the estimate of $T_{w,1}$ closer to 25$^{\circ}$C (or, to be more precise, 
to within the range 13--70$^{\circ}$C, depending on the effective diameter 
of a pentane molecule that one assumes. Adopting a value of 4.4 {\AA}, which
follows from the excluded volume used in the Peng--Robinson equation of
state, yields $T_{w,1} = 38^{\circ}$C.) \cite{bertrand,dobbs}. 

Within this theoretical setting, the observed sequence of wetting transitions  
(`sequential wetting') is brought about by the interplay of short-range and
long-range forces. At low temperatures, both long-range and short-range
interactions inhibit the formation of a wetting layer. At $T_{w,1}$ (the
wetting temperature predicted by Cahn theory based on short-range forces
alone), the short-range forces start to favor a wetting layer
(in a sense that the gain in free energy from having such a layer outweighs 
the cost of creating an additional liquid--vapor interface), the 
long-range forces, however, still act against the formation of a
macroscopically thick film; the result of this interplay is a compromise:
the mesoscopically thick film, which is present in an intermediate wetting
state that has been termed `frustrated-complete wetting' \cite{bertrand}. 
At $T_{w,c}$, the effect of long-range
forces changes its nature from inhibiting to supporting a thick wetting
layer and, consequently, the layer thickness diverges.

This interplay also illustrates why sequential wetting has been observed only 
for such a small number of systems: the window of opportunity for it to occur
is relatively narrow, and, in most cases, the long-range forces will support
wetting before the short-range forces do, so that only a first-order wetting
transition is observed at the temperature at which the short-range forces
start to favor wetting (as in standard Cahn theory).

As already mentioned, the location of the critical-wetting transition
is relatively easily calculated from the variation of the Hamaker constant
as a function of temperature. Within Israelachvili's approximation
\cite{israel}, only the static dielectric constants, the refractive indices,
and a typical absorption frequency of the substances involved are required.

The location of the first-order thin--thick transition, however, is more
difficult to predict: within a Cahn-type theory, knowledge of the so-called
contact energy, which describes the short-range interaction between
substrate and adsorbate (as deduced from measurements of the surface
pressure, for example \cite{ragil2}) is needed. Furthermore, the conventional 
Cahn theory underestimates the (first-order) wetting temperature of alkanes
on water considerably \cite{ragil2} -- only a modification introduced by 
Dobbs \cite{dobbs} allows one to predict $T_{w,1}$ (semi-)quantitatively,
but it requires an effective
diameter of the adsorbate molecule on the assumption that the latter is
spherical \cite{bertrand,dobbs}.

For pentane on water, it has been shown that this kind of theory, when
combined with appropriate expressions for the amplitudes of the
long-distance tails of the long-range
forces \cite{indekeu}, is able to reproduce both wetting-transition
temperatures accurately, $T_{w,1} = 298$ K and $T_{w,c} = 326$ K\@. 
\cite{weiss} Note, however, that, using the same parameters, the short-range 
forces alone (Dobbs' theory) would predict
$T_{w,1} = 311$ K, so the long-range forces are not qualitatively, but, 
to some degree, quantitatively important also for the first-order
transition. 

Another unexpected finding in the experiments on the hexane/brine system,
in addition to the occurrence of sequential wetting, was that the two 
transition temperatures, $T_{w,1}$ and $T_{w,c}$, were shifted {\em in
parallel}\/ as a function of the salt concentration \cite{shahidzadeh}. 
On theoretical grounds,
it had been expected that the (critical) wetting temperature decreases with
increasing salinity of the substrate \cite{richmond}. This behavior 
is indeed observed in the experiments \cite{shahidzadeh}. The first-order 
transition temperature, however, was
expected \cite{shahidzadeh,ragilth} to remain largely unchanged because 
the brine/air(vapor) and 
brine/alkane interfacial tensions vary in a similar fashion as a function
of the salt concentration \cite{aveyard}. It was even hoped that the two
lines of transition temperatures would meet to form a critical endpoint 
\cite{ragilth} (like in a simple model system of sequential wetting 
\cite{weiss2}) -- this expectation, however, was not met: as already
mentioned, the two lines run in parallel and, at first glance, the two
transition temperatures might appear coupled \cite{shahidzadeh,bertrand2}. 
(In contrast to the
situation at $T_{w,c}$, which is necessarily the temperature at which
the Hamaker constant $W$ changes sign, there is nothing peculiar or exceptional
about the behavior of $W$ at $T_{w,1}$, a temperature which is not determined 
by the long-range forces alone. \cite{bertrand,indekeu2}) In an attempt to 
illustrate that $T_{w,1}$ (or
at least $\Delta T_{w,1}$, the shift of $T_{w,1}$ as compared to the case of 
hexane on pure water) is determined by the same dielectric properties as  
$T_{w,c}$, Bertrand {\em et al.}\/ proposed a 
Dzyaloshinskii--Lifshitz--Pitaevskii(DLP)-type theory \cite{dlp} to calculate 
the change in contact energy between substrate and adsorbate
on addition of salt for the system of hexane on brine\cite{bertrand2}. With 
the aid of a Cahn phase portrait (for hexane
on pure water, neglecting long-range forces), the contact-energy difference
is then converted into a shift of the first-order transition temperature.
The DLP-type calculation takes into account the existence of a depletion
layer at the brine/alkane interface, which is almost completely devoid of
ions \cite{aveyard,onsager,massoudi}. The calculation of the free energy 
per unit area 
therefore involves four layers: brine/water/liquid alkane/vapor
\cite{bertrand2}. If the thickness of the water
layer is assumed to be 2 {\AA}, which is consistent with estimates based
on the Onsager--Samaras theory of the interfacial tension of electrolytes
\cite{onsager},
which attributes the ions being repelled from the interface to
electrostatic effects due to image charges, one obtains first-order wetting
temperatures that are in very good agreement with the experimental 
results \cite{bertrand2}. 
Since the thickness $\delta$ does not depend on the Hamaker constant, this
actually just demonstrates that DLP theory works even on very short length 
scales, for which it was not designed originally, but in no way does it
support the supposition that the Hamaker constant might determine $T_{w,1}$ 
as well.

Within Onsager--Samaras theory, however, $\delta$ is a function of the salt
concentration $c_{\rm NaCl}$, as Bertrand {\em et al.}\/ pointed out
\cite{onsager,bertrand2}. If this dependence
is taken into account, the agreement of the calculated $T_{w,1}$ with the
experimental values deteriorates quickly as $c_{\rm NaCl}$ increases. 
Bertrand {\em et al.}\/ speculatively attribute this behavior to the breakdown 
of the underlying Debye--H\"uckel theory for higher concentrations 
\cite{bertrand2}.

In this paper, we take a somewhat different view and develop a theory
that -- also based on DLP theory -- accounts for the modifications of the
long-range forces due to the presence of salt. The system we consider is
similar in spirit to the four-layer structure (brine/water/liquid
alkane/vapor); in order to have a complete description of our system
within the theoretical framework, we also take into account the first
layer of adsorbed hexane molecules, which is denser than the bulk liquid
hexane, as a separate, fifth medium and, thus, deal with a five-layer
structure (brine/water/dense liquid hexane/liquid hexane/vapor). To
estimate the thickness of the layer of pure water, $\delta$, however, we do 
not rely
on Onsager--Samaras theory, which is not valid in the concentration range
of interest here (0.5 mol/L $ \le c_{\rm NaCl} \le $ 2.5 mol/L), but only up 
to $c_{\rm NaCl} \approx 0.15$ mol/L \cite{levin}. Instead, we adopt an idea 
that has 
recently been advanced by Levin and Flores-Mena in order to improve upon
Onsager--Samaras theory for higher salt concentration \cite{levin2}. They
had realized that -- in addition to the effect of an electrostatic repulsion of 
the ions from the interface -- there is a {\em concentration-independent}\/ 
depletion
layer, the thickness of which corresponds to the radius of the {\em
hydrated}\/ ion, which is also about $\delta =  2$ {\AA} for NaCl 
\cite{levin2,conway}. Therefore, the apparent puzzle of having a constant 
thickness of the depletion layer is not due
to the breakdown of Debye--H\"uckel theory (which seems to work fine even
for relatively high concentrations of 1 mol/L in the theory of Levin and 
Flores-Mena \cite{levin2}), but is due to the omission of the effect of the 
hydration-sphere
contribution in Onsager--Samaras theory. This theory had been designed for the
low-concentration regime, in which the depletion-layer thickness is 
dominated by electrostatic effects, whereas it is the hydrodynamic ionic 
radius that sets $\delta$ for the concentration range in which we are 
interested here.

Now we argue that the contact energy in the hexane/brine system is still
the same as in the hexane/pure-water case because, as before, only pure
water is in direct contact with the adsorbed hexane. What changes upon 
the addition of salt is
the long-range interaction between the substrate, now brine, and the adsorbate,
hexane, across the water `layer' of constant thickness $\delta$. Based
on these insights, we will develop a new theory to compute the first-order
wetting temperatures. 

The remainder of the paper is organized as follows: in the next section,
we will explain our approach in detail. The results of our calculations
will be presented in Sec.~III\@. A discussion of
the results and of future tasks in Sec.~IV will conclude the article.

\section{METHODOLOGY AND OUTLINE OF THE THEORY}

This section contains the theoretical framework by means of which we
describe the wetting properties of an alkane (hexane in this case) on pure
water and on brine, respectively. First, we will briefly summarize the 
equation of state
that is used to represent and compute the bulk properties of hexane. 
The fact that hexane and (salt) water are hardly miscible allows us
to use an equation of state for the pure adsorbate and enables us to manage
without an equation of state for the substrate. 
In the second subsection, we present the modified Cahn--Landau model
for the interfacial tensions, which, in turn, determine the wetting 
properties. In that subsection, we will also give a detailed account
of how the long-range field is modified by the presence of salt and
of a depletion layer. For completeness and reproducibility, we also list
the representative equations for the dielectric properties of all five
media involved. A third subsection will contain some of the technical 
details of our calculations.

\subsection{Equation of state for hexane}

As in several previous works on the wetting properties of alkanes
(on aqueous substrates) \cite{ragil,dobbs,ragil2,weiss}, we employ the 
Peng--Robinson
equation of state to describe the thermophysical bulk properties of hexane
\cite{peng}. According to
this equation of state, the pressure is given by:
\begin{equation}
   P = \frac{\rho R T}{1 - b \rho} + \frac{a(T) \rho^2}{1 + b \rho (2 - b
       \rho)},
\end{equation}
where $T$ is the absolute temperature, $R$ the molar gas constant, and $\rho$
the molar density. The excluded volume $b$ is determined from critical
parameters (see below), while the function $a(T)$ also involves the vapor
pressure at the reduced temperature $t = T/T_c = 0.7$. Adopting Peng
and Robinson's recommendations, we use the following prescriptions for the
above-mentioned parameters:
\begin{eqnarray}
   b & = & 0.07780 \, R T_c / P_c , \\
   a(T) & = & a(T_c) \alpha(t, \omega) , \\
   a(T_c) & = & 0.045724 \, (R T_c)^2 / P_c .  
\end{eqnarray}    
In the second equation, $\omega$ is the acentric factor given by
\begin{equation}
   \omega = - 1 - \log_{10}{\left [ P(t=0.7)/P_c \right ]}; 
\end{equation}
the function $\alpha(t, \omega)$ is defined as
\begin{eqnarray}
   \alpha(t, \omega) & = & \left[ 1 + K(\omega) \left( 1 - t^{1/2} 
                             \right) \right]^2 \\
   K(\omega) & = & 0.37464 + 1.54226 \, \omega - 0.26992 \, \omega^2 .
\end{eqnarray} 
The critical and other required parameters for hexane are
\cite{lide,prausnitz}:
\begin{eqnarray}
   T_c & = & 507.7 \, {\rm K} \\
   P_c & = & 3.010 \, {\rm MPa} \\
   \omega & = & 0.296.  
\end{eqnarray}
 
\subsection{Model for the interfacial tension}

Here, we distinguish between two cases and present the theories to describe
the wetting behavior of hexane on pure water and on brine separately.

\subsubsection{Hexane on pure water}

For hexane on pure water, we use a model of the interfacial tension that
is completely analogous to the one employed by Weiss and Widom for pentane
on water \cite{weiss}. This model, in turn, is a combination of Dobbs'
modified Cahn theory \cite{dobbs} and a treatment of the algebraic tails
of the long-range forces as proposed by Indekeu {\em et al.} \cite{indekeu} 
Within this model, the interfacial tension is obtained
by minimizing the free-energy functional with respect to the density profile
of the adsorbate, $\rho(z)$, in conjunction with the appropriate boundary
conditions (see Sec.~II.C for details). The free-energy functional reads:
\begin{eqnarray}
   \gamma[\rho] & = & \gamma_0 + \phi(\rho_0) + \int\limits_{\Delta z}^{\infty} 
                  \left\{ \Delta f(\rho, \rho_{bulk}) + \frac{c}{2}
                  \left( \frac{d \rho}{d z} \right)^2 \right\} dz \nonumber\\
                & & +   \left\{ \Delta f(\rho_0, \rho_{bulk}) + \frac{c}{2} 
                  \left[(\rho_0 - \rho_1) / \Delta z\right]^2\right\}
                  \Delta z \nonumber \\
                & & - \int\limits_{z_c}^{\infty} \left( \frac{a_3}
                  {z^3} + \frac{a_4}{z^4}\right) \rho(z) dz. 
                  \label{intfacten}
\end{eqnarray}
In this model, the $z$-axis is perpendicular to the substrate surface
and the substrate (water) is assumed to occupy the lower half-space, for which
$z<0$, so that the planar water/hexane interface is located at $z=0$.
For $z>0$, $\rho(z)$ denotes the spatially varying density of hexane, while
$\rho_{bulk}$ is the density of either bulk phase, liquid or vapor. The
density of the adsorbate at the substrate surface, $\rho(z=0)$, is denoted
by $\rho_0$.
The first term, $\gamma_0$, represents that part of the interfacial tension
that is due to the self-interaction of the substrate (water/vacuum), but 
since it is only a function of temperature and an additive term
contributing equally to each interfacial tension (except for $\gamma_{lv}$,
which does not involve the interaction with a substrate), it is irrelevant
to the wetting properties of the adsorbate, and no value of $\gamma_0$
needs to be specified. In the second term, $\phi(\rho_0)$ denotes the 
contact energy, which depends only on the surface density of the adsorbate. 
Within the framework of his modified Cahn theory, Dobbs \cite{dobbs} 
determined the contact energy
for several $n$-alkanes on water from experimental data for the surface
pressure, applying a procedure that had been proposed by Ragil {\em et
al.}\/ \cite{ragil2} using standard Cahn theory. For hexane on pure water, we 
find from Fig.\ 2 in Dobbs' paper:
\begin{equation}
   \phi(\rho_0) = \left[ -4.16 \rho_0 b + 2.15 (\rho_0 b)^2 \right] P_c 
                  \lambda. 
\end{equation}     
Here, $\lambda$ is a characteristic length scale of the system (approximately
34 {\AA}) defined by $\lambda = (c/(b^2 P_c))^{1/2}$, where $c$ is the 
influence parameter
(or simply the coefficient of the square-gradient term). By matching to
the experimental surface tension data of $n$-alkanes, Carey {\em et
al.}\/ \cite{carey}
found that $c$ can be estimated from the parameters $a$ and $b$ in the 
Peng--Robinson equation of state for $n$-alkanes of short and medium
chain length (on the basis of a corresponding-states idea). Their relation
reads \cite{carey}:
\begin{equation} \label{sqgrcoeff}
   c = 0.27 N_A^{-2/3} a b^{2/3} + 7.25 \times 10^{-20} {\rm J \, m^5 \, 
       mol^{-2}}
\end{equation}
with $N_A$ being Avogadro's constant.

The third term in Eq.\ (\ref{intfacten}) is the continuum part from
standard Cahn--Landau theory, where $\Delta f$ measures the excess free
energy corresponding to the local density over that of either bulk phase.
It is given by $\Delta f(\rho,\rho_{bulk}) = f(\rho) - \rho \mu_{bulk} +
P_{bulk}$; here, $f = F/V$ is the Helmholtz free-energy density, 
$\mu_{bulk}$ the bulk chemical potential, and $\rho$ the local density of
hexane, $\rho(z)$. The coefficient $c$ is the influence parameter as given
by Eq.~(\ref{sqgrcoeff}).
The integration is to be performed from $\Delta
z$ to infinity in this case, where $\Delta z$ is the `thickness' of the
first layer of adsorbed hexane molecules \cite{dobbs}. Treating the hexane 
molecule as spherical and estimating its diameter $\sigma$ from the 
excluded-volume term in the Peng--Robinson equation of state, one obtains 
$\Delta z = \sigma = 4.4$ \AA
\cite{bertrand}. The first layer of hexane molecules ($0 < z < \Delta z$)
is considered explicitly in a lattice-gas approach and contributes the fourth 
term to Eq.\ (\ref{intfacten}), which is a discretized version of the third. 
The density at a distance $\Delta z$ from the substrate is denoted by $\rho_1$.

The last term contains the long-range field, which -- in a first
approximation -- couples linearly to the density. Following Indekeu {\em et
al.}, we adopt a lower cutoff of $z_c = 0.2 \lambda$, which worked well
for pentane \cite{indekeu,weiss}; for hexane as the adsorbate, we,
therefore, have $z_c \approx 6.8$ {\AA}. This value is also very close
to $1.5 \, \sigma \approx 6.6$ {\AA}, the distance of closest approach
of adsorbate particles that are not in the first layer, but part of the
continuum in the modified Cahn theory. The amplitudes of the first couple
of leading terms in a $1/z$-expansion of the long-range forces, $a_3$ and 
$a_4$, are calculated from DLP theory as follows:       
invoking Israelachvili's approximation \cite{israel}, which amounts to the 
assumption that
all media involved have the same characteristic absorption frequency
$\omega_e$ and that the dielectric spectrum of substance $j$ can be 
represented by $\varepsilon_{j}(i \zeta) = 1 + (n_j^2 -1)/(1 +
(\zeta/\omega_e)^2)$, with $n_j$ being the refractive index and $\zeta$
denoting the frequency, the Hamaker 
constant of a water(1)/liquid hexane(3)/vapor(2) system is given by
\cite{israel}: 
\begin{eqnarray}\label{hamakerconst}
   W & = & \frac{3}{4} k_B T \, \left( \frac{\varepsilon_{3}(0) - 
           \varepsilon_{1}(0)}
           {\varepsilon_{3}(0) + \varepsilon_{1}(0)}\right) \,
           \left( \frac{\varepsilon_{3}(0) - \varepsilon_{2}(0)}
           {\varepsilon_{3}(0) + \varepsilon_{2}(0)} \right) \nonumber\\
     & & + \frac{3 \hbar \omega_e}{8 \sqrt{2}} \frac{
           \left( n_3^2 - n_1^2 \right) \left( n_3^2 - n_2^2 \right)}
           {\left( n_3^2 + n_1^2 \right)^{1/2} \left( n_3^2 + n_2^2
           \right)^{1/2} \left[ \left( n_3^2 + n_1^2 \right)^{1/2} + \left(
           n_3^2 + n_2^2 \right)^{1/2}\right]} \label{hamconst} .
\end{eqnarray}
In this equation, $k_B$ is Boltzmann's constant, while $\hbar$ denotes 
Planck's constant.
The coefficient $a_3$ is related to $W$ by $a_3 = - W/(6 \pi \rho_l)$,
where the denominator should actually contain the difference $\rho_l -
\rho_v$, but in view of the large distance from the bulk critical point
of hexane, $\rho_v$ can safely be neglected \cite{indekeu}.  

To calculate $a_4$, we follow the approach of Bertrand {\em et al.}\/
\cite{bertrandth,bertrandepl} and consider a four-layer structure 
water(1)/dense liquid hexane(4)/liquid
hexane(3)/vapor(2) because, right at the water/hexane interface, there is
a thin layer (of approximately one molecular diameter thickness) of
hexane whose density is higher than the bulk liquid density at
the respective temperature; cf.\ Fig.~1 (a). Since the layer of liquid hexane, 
the thickness of which we denote by $l$, is assumed to be much thicker than 
just one molecular diameter $\sigma$ in the frustrated-complete wetting
state (and certainly so in the complete-wetting state), we expand the full 
free energy per unit area of the four-layer structure, as given by Mahanty and
Ninham \cite{mahanty} and by Parsegian and Ninham \cite{parsegian}, in powers 
of $\sigma/l$
and truncate the resulting series after the
linear term, since $\sigma \ll l$. The Hamaker constant and, therefore,
the expression for $a_3$ remain unchanged by this consideration: the
$l^{-2}$-term gives the Hamaker constant just as in Eq.\ (\ref{hamconst}),
while the $\sigma/l^3$-term results in a coefficient $B$ as
appearing in the following expansion of the long-range part of the
free energy per unit area:
\begin{equation} \label{fexpipol}
   \gamma_{\rm LR}(l,\sigma) = - \frac{W}{12 \pi l^2} + 
                               \frac{B \, \sigma}{12 \pi l^3},
\end{equation} 
where our result for $B$ is given by:
\begin{eqnarray} \label{ourb}
   B & = & - \frac{3}{2} k_B T  \, \left( \frac{\varepsilon_{3}
           (0) \left[ 
           \varepsilon_{1}(0) - \varepsilon_{4}(0)\right] \left[ 
           \varepsilon_{1}(0) + \varepsilon_{4}(0)\right]} 
           {\varepsilon_{4}(0) \left[ 
           \varepsilon_{1}(0) + \varepsilon_{3}(0)\right]^2} \right) \,
           \left(
           \frac{\varepsilon_{3}(0) - \varepsilon_{2}(0)}
           {\varepsilon_{3}(0) + \varepsilon_{2}(0)} \right) \nonumber \\
     & & - \frac{3 \hbar \omega_e}{4 \sqrt{2}} \left( n_3^2 - n_2^2
           \right) \left( n_1^2 - n_4^2 \right) A,    
\end{eqnarray}
with $A$ being:
\begin{eqnarray}
  A  & = & \frac{\left( n_1^2 - n_3^2 \right) \left( n_3^2 - n_4^2 \right)}
           {2 \left( n_1^2 - n_2^2 \right) \left( n_1^2 + n_3^2 \right)^{3/2}
           \left( n_1^2 + n_3^2 - 2 n_4^2\right)} \nonumber \\ 
     &   & + \frac{\sqrt{2} \left( n_1^2 - n_4^2 \right) \left( n_3^2 -
           n_4^2 \right)}{n_4 \left( n_1^2 + n_3^2 - 2 n_4^2\right)^{2}
           \left( n_2^2 + n_3^2 - 2 n_4^2\right)} \nonumber \\ 
     & &   + \frac{\left( n_2^2 - n_3^2 \right) \left( n_1^2 - n_2^2 -
           n_3^2 + n_4^2 \right)}{\left( n_1^2 - n_2^2 \right)^2 
           \left( n_2^2 + n_3^2 \right)^{1/2} \left( n_2^2 + n_3^2 - 2
           n_4^2\right)} \nonumber \\ 
     & &   + \frac{-n_1^6 + n_1^4 \left( n_2^2 + n_3^2 + n_4^2\right) - n_1^2
           \left( n_3^4 + 2 n_2^2 n_4^2 \right)}
           {\left( n_1^2 - n_2^2 \right)^2
           \left( n_1^2 + n_3^2 \right)^{1/2} \left( n_1^2 + n_3^2 - 2
           n_4^2\right)^2} \nonumber \\
     & &   + \frac{ - n_3^2 \left( n_3^4 - 3
           n_3^2 n_4^2 + 2 n_4^4\right) + n_2^2 \left( n_3^4 - 2 
           n_3^2 n_4^2 + 2 n_4^4\right)}
           {\left( n_1^2 - n_2^2 \right)^2 
           \left( n_1^2 + n_3^2 \right)^{1/2} \left( n_1^2 + n_3^2 - 2
           n_4^2\right)^2} .
\end{eqnarray}
We remark that, since $B$ is approximately of the same order 
of magnitude as $W$ ($10^{-20}$ J) far from $T_{w,c}$ (it is actually four
times larger than $W$ at $T = 0^{\circ}$C and ten times larger than $W$ at
$T_{w,1}$), the second term in Eq.~(\ref{fexpipol}) is
indeed typically smaller than the first by a factor of $\sigma/l$. In 
contrast, near $T_{w,c}$, $W$ is negligible and $B$ becomes 
all-important. 

The amplitude $a_4$ is related to $B$ via $a_4 = B \sigma/(4 \pi \rho_l)$, 
where again, $\rho_v$ has been neglected in the denominator since, far from the 
critical point of hexane, $\rho_l \gg \rho_v$. Table I contains the 
representative equations for the
dielectric properties involved in the above expressions. The static
dielectric constant and the refractive index of the hexane layer near the
water surface is estimated using the Clausius--Mossotti and Lorenz--Lorentz
equations, respectively, assuming that the density of hexane is enhanced by 
about 12\% compared to the bulk density of the liquid \cite{ragil2,bertrandth}. 
Numerically, the results for $B$ from the above expression are very close 
to those obtained from the corresponding relation derived by Bertrand {\em
et al.} \cite{bertrandth,bertrandepl}, which, in addition to the truncated 
$\sigma/l$-expansion, 
assumes $\varepsilon_3 \approx \varepsilon_4$ as well as $n_3 \approx n_4$ and
is, therefore, limited to quadratic order in the terms $\Delta_{jk}(i
\zeta)$, where $\Delta_{jk}(i \zeta) = \left( \varepsilon_{j}(i \zeta) - 
\varepsilon_{k}(i \zeta) \right)/
\left( \varepsilon_{j}(i \zeta) + \varepsilon_{k}(i \zeta) \right)$.
This additional approximation does not allow one to recover the 
formally correct result for the Hamaker constant in this 
four-layer calculation. Bertrand's approximate relation 
\cite{bertrandth} reads:
\begin{equation}\label{bertrandb}
   B \approx \frac{3}{2} \, k_B T \Delta_{32}(0) \Delta_{41}(0) + 
       \frac{3}{4 \sqrt{2}} \,
       \hbar \omega_e N(3,2,4,1),
\end{equation}  
where
\begin{equation}
   N(f,g,j,k) = \frac{\left( n_f^2 - n_g^2 \right) \left( n_j^2 - n_k^2
                \right)}{ \left( n_f^2 + n_g^2 \right)^{1/2} \left( n_j^2 +
                n_k^2 \right)^{1/2} \left[ \left( n_f^2 + n_g^2
                \right)^{1/2} + \left( n_j^2 + 
                n_k^2 \right)^{1/2}\right]}.
\end{equation}

In the following, we propose an alternative approach, which will prove 
advantageous in the next subsection concerning the case of hexane on brine.
It, too, is a quadratic 
approximation in the quantities $\Delta_{jk}$ but recovers the exact
limits for $\sigma \to 0$ and $\sigma \to \infty$. According to Mahanty and 
Ninham
\cite{mahanty}, the expression for $\Delta_{3;41}(\sigma)$ in a four-layer
structure, in which medium 3 is of fixed thickness $l$, and media 1 and 3 are 
separated by an intervening layer of medium 4, whose thickness is $\sigma$
(cf.~Fig.\ 1 (a)), 
reads (omitting the frequency-dependence of $\Delta$ for brevity) 
\begin{equation}\label{mahaninhd31}
   \Delta_{3;41}(\sigma) = \frac{\Delta_{34} + \Delta_{41} e^{- \sigma x / l}}
                    {1 + \Delta_{34} \Delta_{41} e^{- \sigma x / l}},
\end{equation}
where $x$ is a dimensionless variable of integration. Note that 
$\Delta_{3;41}(0) = \Delta_{31}$ and $\Delta_{3;41}(\infty) = \Delta_{34}$.
In Bertrand's approximation \cite{bertrandth}, which is conform with the 
expansion proposed by Mahanty and Ninham \cite{mahanty}, this expression 
reduces to 
$\Delta_{3;41}(\sigma) \approx \Delta_{34} + \Delta_{41} e^{- \sigma x / l}$, 
which yields the exact result for $\sigma \to \infty$, but not for $\sigma 
\to 0$. In the limit $\sigma \to 0$, one has $\Delta_{3;41}(0) \approx 
\Delta_{34} +  \Delta_{41}$, which is not exact but correct to order
$\Delta^2$. We now observe that $\Delta_{41} = \Delta_{31}
- \Delta_{34} + O(\Delta^2)$, using Eq.~(\ref{mahaninhd31}) at $\sigma =
0$. Replacing $\Delta_{41}$ by this expression in 
Eq.~(\ref{mahaninhd31}), we obtain $\Delta_{3;41}(\sigma)
\approx \Delta_{34} + \left( \Delta_{31} - \Delta_{34}\right) e^{- 
\sigma x / l}$, which is correct to order $\Delta^2$ {\em and\/} yields the 
exact results in both limits
$\sigma \to 0$ and $\sigma \to \infty$. It, therefore, allows one
to obtain the exact Hamaker constant $W$ in addition to an expression for
$B$ within an approximation to quadratic order in $\Delta_{jk}$, in the
context of an expansion to first order in $\sigma/l$. The latter
expression reads:
\begin{equation}\label{newb}
   B \approx \frac{3}{2} \, k_B T \Delta_{32}(0) \left[ \Delta_{31}(0)
             - \Delta_{34}(0) \right] + 
       \frac{3}{4 \sqrt{2}} \,
       \hbar \omega_e \left[ N(3,2,3,1) - N(3,2,3,4)\right].
\end{equation}   
Numerically, Eqs.\ (\ref{bertrandb}) and (\ref{newb}) give very similar
results; to within a resolution of 0.5 K, there is no detectable difference
in the resulting $T_{w,1}$.

In the actual
process of minimizing the free-energy functional Eq.\ (\ref{intfacten}),
we use representative equations for the amplitudes $a_3$ and $a_4$,
which were obtained by linear regressions to the results of Eqs.\ 
(\ref{hamakerconst}) and (\ref{newb}), and which, for hexane on pure water, 
read:
\begin{eqnarray}
   a_3 & = & \left( 4.681 \times 10^{-23} {\rm J \, K^{-1}} (T - 273.15 \, 
             {\rm K}) - 
             4.460 \times 10^{-21} {\rm J}\right)/\left( 6 \pi \rho_l \right) \\
   a_4 & = & \left( -5.432 \times 10^{-32} {\rm J m \, K^{-1}}(T - 273.15 \, 
             {\rm K}) + 8.987 \times 10^{-30} {\rm J m}\right)/
             \left( 4 \pi \rho_l \right).
\end{eqnarray}
Using the above expressions, we obtain $T_{w,1} = 346$ K and $T_{w,c} = 
369$ K in nearly perfect agreement with the values that were extrapolated 
from the experimental transition temperatures for hexane on brine 
using different concentrations of salt (2.5, 1.5, and 0.5 mol/L) down to
$c_{\rm NaCl} = 0$ mol/L \cite{shahidzadeh}.   

\subsubsection{Hexane on brine}

As already mentioned in the introduction, the fact that ions dissolved
in water carry a hydration sphere around them, which prevents the ionic
centers from approaching the (salt) water/alkane interface 
any closer than the distance set by the radius of this sphere, creates a 
depletion `layer' of  pure water near this interface \cite{levin2}. 
Despite the fact that DLP theory was derived for much larger 
separations (thicker films) than the molecular dimensions we are dealing 
with here, it seems to work reliably even in these extreme cases of very 
short length scales.
With this additional `layer' of pure water present, we now consider a
five-layer structure consisting of brine(1)/water(4)/dense liquid
hexane(5)/liquid hexane(3)/vapor(2); see Fig.~1 (b). In this calculation 
the brine and
vapor phases are semi-infinite slabs, while the water layer is of
thickness $\delta$. The layer of dense liquid hexane, again, has molecular
dimensions set equal to the diameter $\sigma$ of a hexane molecule,
and the film of liquid hexane is of thickness $l$.

The origin of the $z$-axis, which marks the border between substrate and
adsorbate, is now located between the layer of pure water and the molecular
layer of dense liquid hexane. Therefore, the major formal difference between
layers (4) and (5) is that the water layer is part of the substrate ($z < 0$,
but this part is not considered explicitly in the free-energy functional),
while the first layer of adsorbed hexane molecules clearly is part
of the adsorbate ($z>0$). 
In the derivation of Eq.~(\ref{ourb}), it was assumed that $l$ is much
larger than $\sigma$, so, strictly speaking, this approximation is
applicable to the frustrated-complete and complete wetting states,
but not to partial wetting. In this latter state, however, the cutoff $z_c$ 
being larger than $\sigma$ ensures that the substrate--adsorbate (dense
liquid hexane, $z < \sigma$) 
interaction is entirely accounted for by the contact energy.

In the present five-layer calculation, the long-range interaction of the 
substrate (brine) with the adsorbate, not the one between the water layer 
and the adsorbate, however, now involves the additional distance $\delta$, 
the thickness of the intervening water layer. The complete expression
for $\Delta_{3;451}(\delta,\sigma)$ with two intervening layers 4 and 5 of
thickness $\delta$ and $\sigma$, respectively, now reads \cite{mahanty}:
\begin{equation}\label{mahaninhd}
   \Delta_{3;451}(\delta,\sigma) = \frac{\Delta_{35} + \Delta_{5;41}(\delta) 
              e^{- \sigma x / l}}
              {1 + \Delta_{35} \Delta_{5;41}(\delta) e^{- \sigma x / l}},
\end{equation}
where $\Delta_{5;41}(\delta)$ is given by 
\begin{equation}
   \Delta_{5;41}(\delta) = \frac{\Delta_{54} + \Delta_{41}
              e^{- \delta x / l}}
              {1 + \Delta_{54} \Delta_{41} e^{- \delta x / l}}.
\end{equation}
Analogously to what was done for $\Delta_{3;41}(\sigma)$ in the previous
section, $\Delta_{5;41}(\delta)$ is now approximated by $\Delta_{5;41}(\delta)
\approx \Delta_{54} + \left( \Delta_{51} - \Delta_{54} \right) 
e^{- \delta x/l}$, which is correct to order $\Delta^2$ and ensures proper 
behavior in the limits $\delta \to 0$ and $\delta \to \infty$. Thus, we have
\begin{equation}
   \Delta_{3;451}(\delta,\sigma) \approx \frac{\Delta_{35} +
      \left[ \Delta_{54} + \left( \Delta_{51} - \Delta_{54} \right) 
      e^{- \delta x/l} \right] e^{- \sigma x / l}}
      {1 + \Delta_{35} \left[ \Delta_{54} + \left( \Delta_{51} -
      \Delta_{54} \right) e^{- \delta x/l} \right] e^{- \sigma x / l}}.
\end{equation} 
Taking the limits $\delta \to 0$ and $\sigma \to 0$ at the same time leads
us to the identity $\Delta_{3;451}(0,0) = \left(\Delta_{35} + \Delta_{51}
\right)/\left( 1  + \Delta_{35} \Delta_{51} \right) = \Delta_{31}$, from 
which we obtain -- 
to quadratic order in $\Delta_{jk}$ -- the approximation $\Delta_{51}
\approx \Delta_{31} - \Delta_{35}$. Therefore, $\Delta_{3;451}(\delta,\sigma)$
now, again to quadratic order, becomes
\begin{equation} \label{del31jos}
   \Delta_{3;451}(\delta,\sigma) \approx \Delta_{35} + \Delta_{54} \,
              e^{- \sigma x / l} + \left( \Delta_{31} - \Delta_{35} -
              \Delta_{54} \right) e^{- \left(\sigma + \delta \right) x / l}.
\end{equation}
Note that this is exact in the two limits $\sigma \to \infty$ and $\sigma =
\delta = 0$.
From here, we proceed by considering the limit $\delta \to \infty$, for
which $\Delta_{3;451}(\delta,\sigma)$ must transform into
$\Delta_{3;54}(\sigma)$. The resulting
approximation for the remaining four-layer structure (4/5/3/2) is 
$\Delta_{3;54}(\sigma) \approx \Delta_{35} + \Delta_{54} e^{- \sigma x / l}$,
which, in the limit $\sigma \to 0$, yields $\Delta_{54} \approx 
\Delta_{34} - \Delta_{35}$. Substituting this into Eq.\ (\ref{del31jos}),
we arrive at 
\begin{equation} \label{del31jos2}
   \Delta_{3;451}(\delta,\sigma) \approx \Delta_{35} + \left( \Delta_{34} -
              \Delta_{35} \right) e^{- \sigma x / l} + \left( \Delta_{31} - 
              \Delta_{34} \right) e^{- \left(\sigma + \delta \right) x / l},
\end{equation}
which is the only approximation correct to quadratic order in $\Delta_{jk}$ 
that describes all limits $\delta \to 0, \infty$ and $\sigma \to 0, \infty$
exactly. Incidentally, note that this approximation leads to expressions
for the free energy per unit area in which the contributions proportional
to $\Delta_{32} \Delta_{35} \, l^{-2}$, $ \Delta_{32} \left( \Delta_{34} -
\Delta_{35} \right)(l + \sigma)^{-2}$ and $\Delta_{32} \left( \Delta_{31} -
\Delta_{34} \right) (l + \sigma + \delta)^{-2}$ are easier
to interpret than in the textbook expressions, such as, e.g.\ Eq.~(5.8)
in the monograph by Mahanty and Ninham \cite{mahanty}. Within this 
approximation scheme, the free energy per unit area arising from long-range
interactions is
\begin{eqnarray} \label{freeenjos}
   \gamma_{\rm LR}(l;\sigma,\delta) & = & 
           - \frac{k_B T}{8 \pi l^2} \sum^{\infty}_{n=0}{'} \,
           \left[ \Delta_{32} \Delta_{35} + \frac{1}{\left( 1 + 
           \sigma/l\right)^2} \Delta_{32} \left( \Delta_{34} - \Delta_{35} 
           \right) \right. \nonumber \\
     &  & \left. + \frac{1}{\left( 1 +  \sigma/l + \delta/l\right)^2} 
           \Delta_{32} \left( \Delta_{31} - \Delta_{34} 
           \right) \right],
\end{eqnarray} 
where the $\Delta_{jk}$ depend on imaginary frequencies $i \zeta_n$, with
$\zeta_n$ given by $\zeta_n = 2 \pi n k_B T / \hbar$, and the contributions
of these frequencies are to be summed up \cite{mahanty}. The prime on the 
summation symbol indicates that the term corresponding to $n = 0$ is to be 
multiplied by a factor of $1/2$. 
Thus, the presence of two intervening layers gives rise to a free energy
per unit area that involves contributions proportional to $1/(l + \delta)^2$ 
and to $1/(l + \delta + \sigma)^2$, respectively. Now, unlike for
$\sigma/l$, we will not expand into powers
of $\delta/l$ and truncate after the linear term (since this approximation
would be valid only for frustrated-complete wetting and for complete
wetting) in order to keep the following physically plausible
picture: while the terms involving $a_3$ and $a_4$ concern the interaction
of the adsorbate with the topmost layer of the substrate (pure water), the 
tails of the long-range field exerted on the adsorbate by brine will be of 
the form $a_3'/(z + \delta)^3$ and $a_4'/(z + \delta)^4$. 
Expanding the above equation to linear order in terms of $\sigma/l$ yields
\begin{eqnarray} \label{freeenjos2}
   \gamma_{\rm LR}(l;\sigma,\delta) & \approx & - \frac{k_B T}{8 \pi l^2} 
           \sum^{\infty}_{n=0}{'} \,
           \left[ \Delta_{32} \Delta_{34} - \frac{2 \sigma}{l} \Delta_{32} 
           \left(
           \Delta_{34} - \Delta_{35} \right) \right.\nonumber \\
     & &  \left. + \frac{1}{\left( 1 +
           \delta/l\right)^2}  \Delta_{32} \left( \Delta_{31} - \Delta_{34}
           \right) - \frac{2 \sigma}{l \left( 1 +
           \delta/l\right)^3} \Delta_{32} \left( \Delta_{31} - \Delta_{34}
           \right) \right] \\
     & = &  - \frac{W}{12 \pi l^2} + \frac{B \, \sigma}{12 \pi l^3} -
          \frac{W'}{12 \pi (l + \delta)^2} + \frac{B' \, \sigma}{12 \pi (l +
          \delta)^3} .
\end{eqnarray}  
This equation is seen to consist of four contributions:
the first and the second term result in the familiar expressions for 
$W$ and $B$, respectively, as given by Eqs.~(\ref{hamakerconst}) and
(\ref{newb}). The third and fourth terms, involving $W'$ and $B'$,
however, act over a distance ($l + \delta$) and, thus, represent a
qualitatively new contribution to the long-range interaction free energy. 
In sum, we obtain the following free-energy functional, which, after 
minimization, will give the interfacial tensions:
\begin{eqnarray} \label{complfreeen}
   \gamma[\rho] & = & \gamma_0 + \phi(\rho_0) + \int\limits_{\Delta
                  z}^{\infty}
                  \left\{ \Delta f(\rho, \rho_{bulk}) + \frac{c}{2}
                  \left( \frac{d \rho}{d z} \right)^2 \right\} dz \nonumber
                  \\
                & & +\left\{ \Delta f(\rho_0, \rho_{bulk}) + \frac{c}{2}
                  \left[(\rho_0 - \rho_1) / \Delta z\right]^2\right\}
                  \Delta z \nonumber \\
                & & - \int\limits_{z_c}^{\infty} \left( \frac{a_3}
                  {z^3} + \frac{a_4}{z^4}\right) \rho(z) dz
                  - \int\limits_{z_c}^{\infty} \left(\frac{a_3'} {(z +
                  \delta)^3} + \frac{a_4'}{(z + \delta)^4}\right) \rho(z) dz
                  \nonumber \\
                & & - \frac{a_3'}{\delta^3} \rho_0 \Delta z.
                  \label{intfactensalt}
\end{eqnarray}
Analogously to the definition of the unprimed coefficients, the primed ones
are given by $a_3' =- W'/ 6 \pi \rho_l$ and $a_4' =  B' \sigma/ 4 \pi \rho_l$,
so we now have:
\begin{eqnarray}
   W & \approx & \frac{3}{4} \, k_B T \Delta_{32}(0) \Delta_{34}(0)  + 
       \frac{3}{8 \sqrt{2}} \, \hbar \omega_e N(3,2,3,4), \\
   B & \approx & \frac{3}{2} \, k_B T \Delta_{32}(0) \left[ 
                 \Delta_{34}(0) - \Delta_{35}(0) \right] \nonumber \\
     & &  + \frac{3}{4 \sqrt{2}} \,
       \hbar \omega_e \left[ N(3,2,3,4) - N(3,2,3,5)\right], \\
   W' & \approx & \frac{3}{4} \, k_B T \Delta_{32}(0) \left[ \Delta_{31}(0)
                  - \Delta_{34}(0)\right]  \nonumber\\
      & & + 
       \frac{3}{8 \sqrt{2}} \, \hbar \omega_e \left[ N(3,2,3,1) - N(3,2,3,4)
       \right], \label{a3prime} \\
   B' & = & 2 \, W' . \label{a4prime}
\end{eqnarray}
Note that for the critical transition, from the frustrated-complete wetting
state to complete wetting, $l$ is very much larger than $\delta$ and the
Hamaker constant, which changes sign at this transition, is given
by $W + W'$.  
In Table II we have compiled representative equations for $a_3'$ and $a_4'$ as
a function of temperature for a few selected concentrations of salt.
For the primed terms, involving $a_3'$ and $a_4'$, respectively, we use
the same cutoff $z_c$ as for the unprimed terms. The actual value of
$z_c$ is very close to $1.5 \, \sigma$, the distance of closest approach
of particles in the region which is treated as a continuum in the modified
Cahn theory employed here. To account for the interaction between brine
and the first layer of adsorbed hexane molecules across the water layer
of thickness $\delta$, we include the last term in Eq.\
(\ref{intfactensalt}). Therefore, the only unknown parameter
introduced by extending the theory from having brine instead of water 
as a substrate is $\delta$. It is this last term in Eq.\
(\ref{intfactensalt}) that, numerically, accounts for the major difference 
between having water and having brine as the substrate. The sensitivity
of our theory to the actual value of $\delta$, which will be discussed in 
detail in the next section, can be traced back to this term.

\subsection{Technical details}

Given the appropriate free-energy (per unit area) functional, Eq.\ 
(\ref{intfacten}) or Eq.\ (\ref{intfactensalt}), respectively, we proceed 
to calculate the different interfacial tensions relevant to the wetting
behavior. For the critical transition, the three interfacial tensions
between substrate (water or brine) and liquid hexane ($\gamma_{sl}$), 
substrate and hexane vapor ($\gamma_{sv}$), and between liquid hexane and 
the vapor phase ($\gamma_{lv}$)
matter. According to Young's equation, complete wetting is reached once
$\gamma_{sv} = \gamma_{sl} + \gamma_{lv}$ (Antonow's rule). In a sequential
wetting scenario, however, the surface free energy per unit area for
the frustrated-complete wetting state, $\gamma_{sv}$, will be extremely 
close to the sum $\gamma_{sl} + \gamma_{lv}$ \cite{weiss}, so that it
becomes very difficult to locate the critical wetting transition exactly
via this route. Fortunately, for long-range critical wetting, $T_{w,c}$ is 
simply determined by the
temperature at which the Hamaker constant changes sign, which allows one
to calculate $T_{w,c}$ very easily without any numerical minimization.
For the first-order transition, we have to
compute the free energies of an $sv$-interface with a mesoscopically thick
film and of an $sv$-interface without such a layer. Only one of these two
configurations will be stable, the other one just metastable, except right 
at $T_{w,1}$, where both are equal in free energy per unit area.

To obtain the four different interfacial tensions, we prescribe simple
initial trial profiles that resemble the final structure and then minimize
the free energy of the system under the respective boundary conditions (see
below) using a conjugate-gradient method. All calculations were done using
2000 points distributed evenly over a distance of 200 {\AA}\@. Several
initial thicknesses of the mesoscopically thick film (60-150 {\AA}) were
tried, but no difference with respect to $T_{w,1}$ could be detected.

The boundary conditions are $\rho(0) = \rho_l, \rho(\infty) = \rho_v$ for
the `free' liquid--vapor interface and $\rho(\infty) = \rho_l$ or $\rho(\infty)
= \rho_v$ for the $sl$ and $sv$-cases, respectively, while $\rho(0)$ is to be
obtained in the process of minimizing the free energy of the whole system
in these cases (natural boundary condition \cite{sagan}).

Note that the long-range terms only contribute significantly
to $\gamma_{sv}$ if $l > z_c$, i.e.\ for the frustrated-complete and
complete wetting states, but not for partial wetting. The long-range
interaction between brine and the first-layer of adsorbed alkane
molecules (last term in Eq.\ (\ref{complfreeen})) is very important and, 
therefore, taken into account in all wetting states. 
         
\section{RESULTS}
 
The main results of our calculations are shown in Fig.~2, which contains
experimental \cite{shahidzadeh} and theoretical first-order and
critical-wetting transition temperatures for hexane on brine as a function
of the salt concentration. The open circles are the two critical-transition
temperatures as determined experimentally for $c_{\rm NaCl} = 1.5$ mol/L
and $c_{\rm NaCl} = 2.5$ mol/L\@. The solid line represents the loci
where the Hamaker constant, as computed from Eq.\ (\ref{hamconst}) using the
equations listed in Table I, changes sign. The agreement with the
experimental data is very good \cite{shahidzadeh}.
The open diamonds denote the three first-order transition temperatures 
that were obtained experimentally for $c_{\rm NaCl} = 0.5, 1.5$, and 
2.5 mol/L\@ \cite{shahidzadeh}. 
Linear extrapolation down to $c_{\rm NaCl} = 0$ mol/L yields a first-order 
transition temperature of 73$^{\circ}$C for hexane on pure water
\cite{shahidzadeh}. The filled squares (connected by a dashed line
as a guide to the eye) are the theoretical values of $T_{w,1}$ as computed
from the theory outlined in Sec.~II.B.2 using a thickness of the
depletion layer of $\delta = 1.9$ {\AA}\@. As can be seen in Fig.\ 2, the 
agreement with experiments is as good as for the critical-wetting
transition temperatures, but a slight bend in the theoretical
curve, which leads to somewhat larger deviations at higher salt 
concentrations, is visible. For $c_{\rm NaCl} = 0$ mol/L, the
first-order transition temperature is calculated according to the
specifications given in Sec.~II.B.1. In this case, the agreement with the 
extrapolated `experimental' value of $T_{w,1} = 346$ K is perfect. 

Naturally, the question of the thickness of the depletion layer does not arise 
for $c_{\rm NaCl} = 0$ mol/L; for $c_{\rm NaCl} >0$ mol/L, however, it does. 
In their theory of the increase of the surface tension of water on addition
of a strong electrolyte, like NaCl, for example, Levin and Flores-Mena 
\cite{levin2} used the value of $\delta =
2.125$ {\AA}, which led to a good description of the experimental
surface-tension data given by Matubayasi {\em et al.}\/ \cite{matubayasi}
for aqueous solutions of this salt.
The theory of Levin and Flores-Mena, however, is quite insensitive to
the actual value of $\delta$: as long as $\delta$ is of the order of 2 {\AA},
it describes the experimental data quite well, as can be seen in Fig.~3
(in which $\delta$ ranges from 1.75 to 2.5 {\AA}).

In sharp contrast to this behavior, the first-order transition temperature
is rather sensitive to small changes of $\delta$. In Fig.\ 4, we have 
compiled the values of $T_{w,1}(c_{\rm NaCl})$ for $\delta = 1.75, 1.9,
2.125$,
and 2.5 {\AA}\@. While these thicknesses are still suitable for describing
the surface tension of NaCl solutions (cf.\ Fig.\ 3), the predictions of 
$T_{w,1}$ are not satisfactory, except for $\delta \approx 1.9$ {\AA}. 
(Note, however, that because of the curvature of $T_{w,1}(c_{\rm NaCl})$,
choosing $\delta = 1.75$ {\AA} yields good agreement
at higher salt concentrations, e.g.\ for $c_{\rm NaCl}$ = 2.5 mol/L.) In 
fact, if one chooses $\delta$ large enough, one may even
create -- purely hypothetically, of course -- the elusive critical
endpoint, where $T_{w,1}$ coincides with $T_{w,c}$.
 
For comparison, we also compute $\delta$ from Onsager--Samaras
theory as Bertrand {\em et al.} did \cite{bertrand2,bertrandth}; the
prediction of $T_{w,1}(c_{\rm NaCl})$ within our approach (see open diamonds
in Fig.\ 4) shows the same downwards-bend as in their theory. 
The disagreement of calculated and experimental values of $T_{w,1}$ is
therefore mainly due to the inapplicability of Onsager--Samaras theory, which 
predicts a varying $\delta$ because it neglects the hydration sphere and
bases the estimate of $\delta$ solely on electrostatic considerations. Doing
so is correct for very low concentrations of salt, but certainly not for the
concentrations present in the experiments on sequential wetting. In 
conclusion, what causes the 
disagreement is {\em not}\/ the breakdown of Debye--H\"uckel theory, as
Bertrand {\em et al.} proposed \cite{bertrand2}, even though this
theory is definitely not applicable for the {\em bulk}\/ properties of a salt
solution in this concentration range. As Levin and Flores-Mena argue
\cite{levin2}, the 
reason why a theory of the surface tension based on Debye--H\"uckel theory
still works for these relatively high salt concentrations is that, in the
canonical-ensemble approach introduced by Levin \cite{levin}, the increase
of the interfacial tension is given by the {\em difference}\/ between the 
bulk free energy of a homogeneous system and that of an inhomogeneous 
system which features the liquid-vapor(air) interface. Therefore, most bulk 
effects simply cancel.       

\section{DISCUSSION AND CONCLUSION}

In the preceding sections, we have demonstrated that a DLP-style theory
\cite{dlp} combined with Israelachvili's approximations \cite{israel} and a 
modern account of the surface tension of electrolytes \cite{levin2} is able to 
describe not only the
critical-wetting transition temperatures, but also the notoriously more
difficult to predict first-order transition temperatures in the 
sequential-wetting scenario of hexane on brine.

A crucial step in the description of the first-order transition temperature
is the determination of the thickness $\delta$ of the depletion layer, a
thin film 
of pure water that forms near the brine/alkane interface. The two attempts
made up to now to describe the wetting temperatures in the sequential wetting 
of hexane on brine rely on a multi-layer (four or five layers, respectively) 
DLP-type
calculation involving layers of brine, water, dense liquid hexane in our
case, liquid hexane, and vapor. In the first approach by 
Bertrand {\em et al.} \cite{bertrand2,bertrandth}, the free energy per unit 
area of such a four-layer structure is
calculated, identified with the shift in contact energy caused by adding
the salt, and subsequently converted into a shift of the first-order
transition temperature as compared to the first-order {\em wetting}\/
transition temperature of hexane on pure water in a conventional Cahn
theory using a phase-portrait technique. The values of $T_{w,1}$ resulting
from this approach were shown to depend sensitively on the value of $\delta$
\cite{bertrand2}. 
Excellent agreement with the experimental data was found for $\delta 
\approx 2$ {\AA}\@.

Similarly, the predictions of $T_{w,1}$ within our approach presented in
Sec.~II.\ are in very good agreement with the experimental results if we
choose $\delta$ to be approximately 1.9 {\AA}. Interestingly, our theory is 
as sensitive
towards changes in $\delta$ as the one by Bertrand {\em et al.} (cf.\
Fig.~4). Therefore, both treatments have descriptive, but hardly any
predictive power. We can conclude that the thickness of the layer of pure
water is 1.8 -- 2 {\AA}, but we would not be able to predict this thickness
to within comparable accuracy for a different salt. In particular, there
is {\em a priori}\/ no reason why, for a different salt, the graphs of
$T_{w,1}(c_{\rm NaCl})$ and $T_{w,c}(c_{\rm NaCl})$ should be parallel; 
the two lines might actually meet in a critical endpoint. In view of the
missing predictive power of the approach, i.e.\ without knowing 
$\delta$ beforehand, however, there
is no point in calculating the two lines of transition temperatures 
for another salt at the current state of theory. 
It is worthwhile noting
that the value of 1.8 -- 2 {\AA} for the hydrodynamic radius of Na$^+$
and Cl$^-$ is consistent with values determined in alternative
ways \cite{conway} and reasonably well suited for describing the surface 
tension of NaCl solutions within the theory of Levin and Flores-Mena (cf.\ 
Fig.~3).

The major differences between the two approaches presented up to now,
by Bertrand {\em et al.} and by us, are, firstly, in the way in which the 
changes due to the addition of salt are attributed to short-range and 
long-range forces, respectively, and, more importantly, in the reasoning why 
the depletion-layer thickness $\delta$ should be constant. With respect to
the former, our approach is to leave the contact energy (clearly
a short-range interaction) unaltered as compared to the case of hexane on
pure water since, as before, due to the presence of the depletion layer of 
thickness $\delta$, only pure water is in direct contact with the adsorbate.
We also retain the long-range forces between the layer of pure water
and hexane; however, there are new terms in the long-range field which 
describe the 
interaction between the brine phase and the adsorbate across the `layer'
of pure water of thickness $\delta$. All contributions to the free-energy
functional in Eq.~(\ref{intfactensalt}) are calculated at the actually 
relevant temperature. Thus, our approach enables us to obtain 
absolute values of $T_{w,c}$ and $T_{w,1}$ if $\delta$ is determined from 
independent measurements of the hydrodynamic radius or deduced by matching 
to the experimental (wetting) transition temperatures.

The second and main difference of our treatment to the one proposed by Bertrand
{\em et al.} is the justification of having a constant, i.e.\ 
salt-concentration independent, depletion-layer thickness $\delta$. Bertrand 
{\em et al.} attribute the existence of this layer to electrostatic image
charges that repel the ions from the brine/alkane interface. Based on
this mechanism, Onsager and Samaras \cite{onsager} had developed their
theory of the increase of the surface tension of water on the addition of
salt. They succeeded in deriving the correct limiting law for low salt
concentrations and -- as a by-product of their theory -- gave an expression
that relates the salt concentration to an {\em equivalent}\/ depletion-layer
thickness $\delta$. The original Onsager--Samaras theory actually calculates
the concentration profile and predicts an exponentially rapid approach
of the local salt concentration to its bulk value as one moves away from the 
interface.
By integration over this profile, one can compute the deficiency of ions
near the interface. This value can, then, on the assumption that there is
a layer which is {\em completely}\/ devoid of ions (but, according to
Onsager and Samaras, this should not be the case), be converted into an
equivalent thickness of such a layer. It is this layer thickness that 
Bertrand {\em  et al.} identify with $\delta$ in their approach.
Within Onsager--Samaras theory, however, this thickness is predicted to
decrease slowly (logarithmically) but significantly with increasing salt 
concentration, so it is 
clear that the layer thickness cannot remain constant in the approach
of Bertrand {\em  et al.} \cite{bertrand2}. It is noteworthy that the 
equivalent 
depletion-layer thickness is only a by-product of the Onsager--Samaras
theory and that it is not actually used to compute the excess surface 
tension of the electrolyte \cite{onsager}.      

Our approach, in contrast, is based on the recent theory of Levin and
Flores-Mena who argue that at higher salt concentrations ($c_{\rm NaCl} >
0.15$ mol/L and, therefore, clearly in the range of salinities present in the 
sequential-wetting experiments) Onsager--Samaras theory neglects the
`intrinsic' depletion layer that is formed by the hydration sphere of
the ions (at low concentrations, this contribution is negligible because
the electrostatic repulsion creates a much thicker layer). In this picture,
it becomes much more transparent why $\delta$ is independent of the salt
concentration (as long as there are enough water molecules for each ion
to have a complete hydration sphere) and about 2 {\AA} in size. The local
salt concentration will increase very rapidly to its bulk value in the
region beyond
the hydration-determined depletion layer thickness because electrostatic
shielding is extremely efficient at high salt concentrations. 
 
In conclusion, the theory presented in this article -- as well as the one by
Bertrand {\em et al.}\/ \cite{bertrand2} -- is able to describe the first-order 
transition temperature in the sequential wetting of hexane on brine if the
thickness of the depletion layer is chosen to be about 1.9 {\AA}\@. The 
predictive capacity of both approaches is, unfortunately, much lower. If one 
had a more precise {\em a priori}\/ knowledge of the hydrodynamic radii of the
ions, 
$T_{w,1}$ and $T_{w,c}$ could be predicted for solutions of different
salts, which would be a major advantage because the actual experiments
are very time-consuming \cite{shahidzadeh}. Such predictions might help
identify suitable candidates of solutes for producing a 
critical endpoint in the wetting phase diagram, which is
of particular interest \cite{shahidzadeh,weiss2,bonn}.

\section*{ACKNOWLEDGMENTS}

We would like to thank E.\ Bertrand and Y.\ Levin for very helpful 
discussions. V.C.W.\ is a visiting postdoctoral fellow of the `Fonds voor
Wetenschappelijk Onderzoek (F.W.O.) -- Vlaanderen' and gratefully acknowledges 
financial support from this organization. Our research has in part been
supported by the European Community TMR Research Network
`Foam Stability and Wetting' under contract no.\ FMRX-CT98-0171.

\clearpage
\newpage

\begin{table}[t]
\caption{Representative equations for the dielectric properties of
water, brine, liquid hexane, and vapor. In addition to the relative static
dielectric permittivity $\varepsilon(0)$ and the refractive index $n$, a
characteristic absorption frequency common to all four media of
$\nu_e = \omega_e/(2 \pi) = 3 \times 10^{15}$ s$^{-1}$, which is in the
UV range, has been used. $T$ denotes the absolute temperature and $c_{\rm
NaCl}$ the salt concentration in the brine phase.} 
\label{tab1}
\begin{center}
\begin{tabular} {ll} \hline \hline
   medium & relative static dielectric permittivity $\varepsilon(0)$ \\ \hline 
   water\protect\cite{lide} & $249.21 - 0.79069 \, {\rm K^{-1}} \, T + 
           0.72997 \times 10^{-3} \, {\rm K^{-2}} \, T^2$ \\ 
   brine\protect\cite{hasted} & $\varepsilon_{\rm water}(0) - 11 \, {\rm L \,
           mol^{-1}} \, c_{\rm NaCl}$ \\
   hexane\protect\cite{weast} & $1.890 - 0.00155 \, {\rm
           K^{-1}} \, (T - 293.15 \, {\rm K})$ \\
   vapor & $\approx 1$ \\ \hline
   medium & refractive index $n$ \\ \hline
   water\protect\cite{ownfit1} & $1.33436 - 1.50585 \times 10^{-5}  \, {\rm
           K^{-1}}\, (T - 273.15 \, {\rm K}) -  
           1.94586 \times 10^{-6}  \, {\rm K^{-2}}$ \\
           & $ \times \, (T - 273.15 \, {\rm K})^2 + 5.23889 
           \times 10^{-9} \, {\rm K^{-3}} \, (T - 273.15 \, {\rm K})^3$ \\
   brine\protect\cite{ownfit2} & $n_{\rm water} + 0.00918 \, 
           {\rm L \, mol^{-1}}\, c_{\rm NaCl}$\\
   hexane\protect\cite{ragilth} & $n_{\rm water} + 0.049 - 0.0004 \, {\rm
           K^{-1}} \, (T - 273.15 \, {\rm K})$\\
   vapor & $\approx 1$ \\
   \hline \hline
\end{tabular}
\end{center}
\end{table}

\begin{table}[t]
\caption{Representative equations for the coefficients $a_3'$ and $a_4'$, 
which are given by $a_3' = -W'/(6 \pi \rho_l)$ and $a_4' = B' \sigma/(4 \pi
\rho_l)$ as calculated from Eqs.~(\protect\ref{a3prime})
and (\protect\ref{a4prime}), respectively,
using the data listed in Table I, for various concentrations of salt. $W'$ 
and $B'$ are the quantities actually listed in the table; $W'$ is
represented in the form $W' = a_{W'} + b_{W'} \, (T -
273.15 \, {\rm K}) + c_{W'} \, (T - 273.15 \, {\rm K})^2 $, where $T$ is the
absolute temperature, and $B'$ is simply given by $B' = 2 \, W'$.}
\label{tab2}
\begin{center}
\begin{tabular} {cccc} \hline \hline
   $c_{\rm NaCl}$ [mol L$^{-1}$] & \multicolumn{3}{c}{$W'$ [J]} \\ 
    & $a_{W'}$ [J] & $b_{W'}$ [J K$^{-1}$]& $c_{W'}$ [J K$^{-2}$] \\ \hline
   0 & 0 & 0 & 0\\ 
   0.5 & $-5.064 \times 10^{-22}$ & $5.329 \times 10^{-25}$ & $1.559 
         \times 10^{-27}$ \\
   1.0 & $-1.012 \times 10^{-21}$ & $1.069 \times 10^{-24}$ & $3.207
         \times 10^{-27}$ \\
   1.5 & $-1.516 \times 10^{-21}$ & $1.609 \times 10^{-24}$ & $4.998
         \times 10^{-27}$ \\
   2.0 & $-2.018 \times 10^{-21}$ & $2.150 \times 10^{-24}$ & $7.033
         \times 10^{-27}$ \\
   2.5 & $-2.519 \times 10^{-21}$ & $2.689 \times 10^{-24}$ & $9.509
         \times 10^{-27}$ \\ \hline \hline
\end{tabular}
\end{center}
\end{table}

\clearpage
\newpage
\begin{figure}
\center{\includegraphics[width=7.5cm,angle=0]{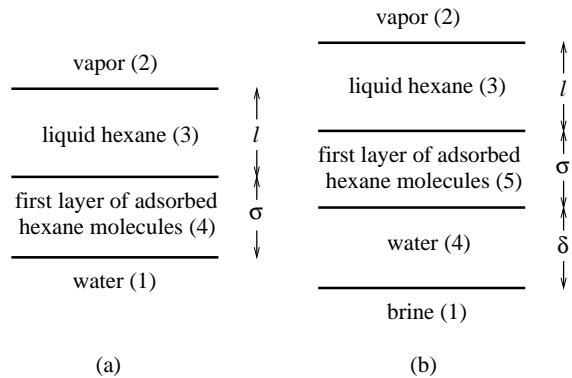}}
\vspace{1cm}
\caption{Part (a) shows the geometry of the four-layer structure
water(1)/dense
liquid hexane(4)/liquid hexane(3)/vapor(2) that is considered for the
wetting of hexane on pure water. Part (b) extends this picture for
the wetting of hexane on brine and introduces a fifth layer in a
structure of the following type: brine(1)/water(4)/dense liquid
hexane(5)/liquid hexane(3)/vapor(2). The film thicknesses of the
various layers (4, 5, 3) are labeled $\delta$, $\sigma$, and $l$,
respectively.}
\end{figure}

\begin{figure}
\center{\includegraphics[width=7cm,angle=-90]{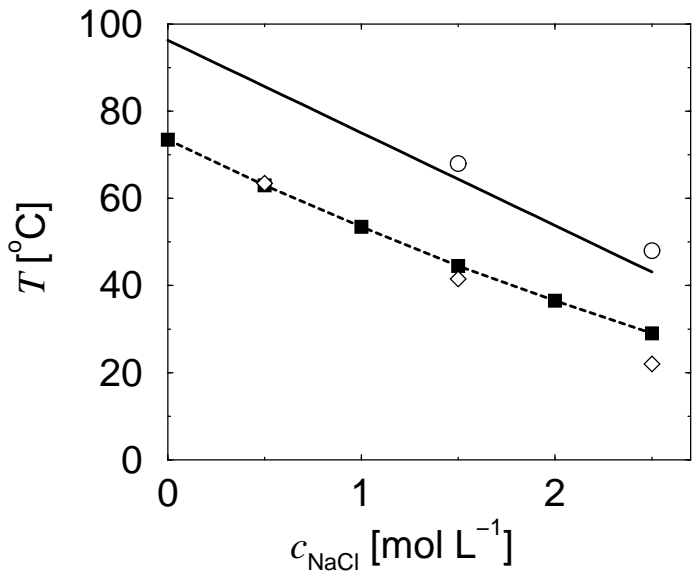}}
\vspace{1cm}
\caption{First-order and critical-wetting transition temperatures for hexane
on brine as a function of the salt concentration. The open circles denote
the experimental critical-wetting temperatures \protect\cite{shahidzadeh},
while the continuous line marks the theoretical prediction for this
transition, as obtained from the temperature at which the Hamaker constant
changes sign. The open diamonds represent the experimental first-order
transition temperatures \protect\cite{shahidzadeh}, and the filled squares,
connected by the dashed line to guide the eye, are our theoretical
predictions using a depletion-layer thickness of $\delta = 1.9$ {\AA}\@.}
\end{figure}

\begin{figure}
\center{\includegraphics[width=7cm,angle=-90]{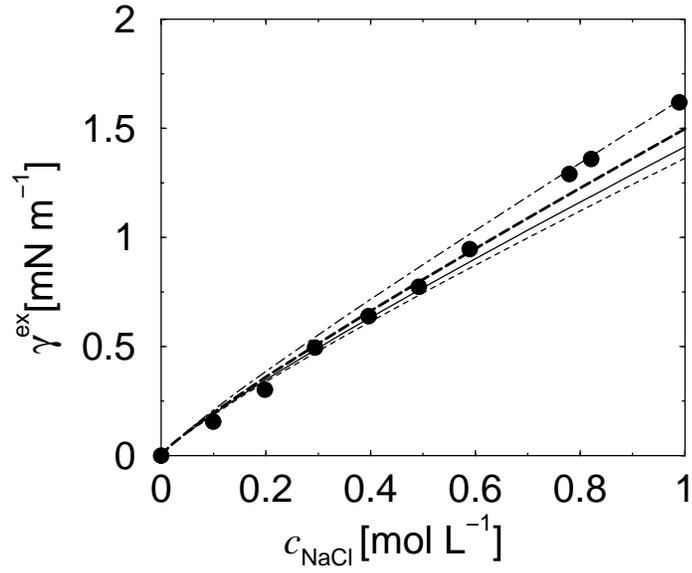}}
\vspace{1cm}
\caption{Excess surface tension of a 1-1 electrolyte, e.g.\ NaCl, over that
of pure
water as a function of the salt concentration. The filled circles represent
the experimental data for NaCl at 25$^{\circ}$C obtained by Matubayasi {\em
et al.}\protect\cite{matubayasi} The lines are results of calculations
based on the theory of Levin and Flores-Mena \protect\cite{levin2} using
different values for the hydrodynamic radius $\delta$ of the ions; from
top to bottom: 2.5 {\AA} (dash-dotted), 2.125 {\AA} (thick long-dashed),
1.9 {\AA} (continuous), 1.75 {\AA} (thin dashed).}
\end{figure}

\begin{figure}
\center{\includegraphics[width=7cm,angle=-90]{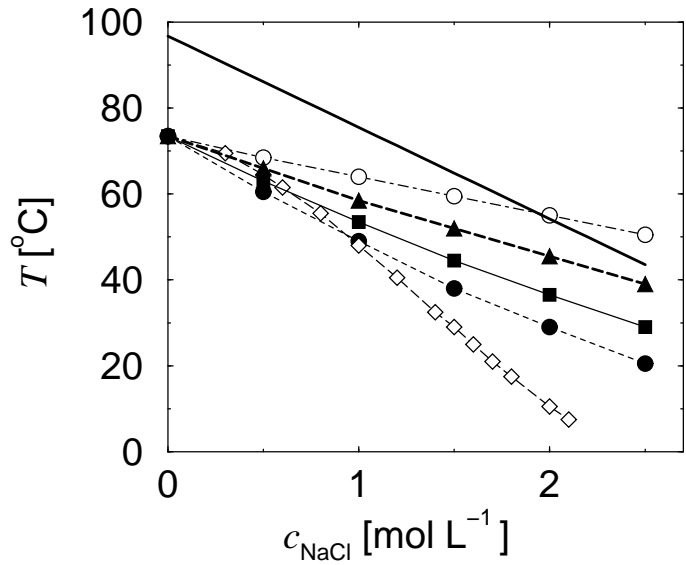}}
\vspace{1cm}
\caption{Predictions of the first-order transition temperatures in the
sequential wetting of hexane on brine for various values of the
depletion-layer thickness $\delta$; from top to bottom: 2.5 {\AA}
(open circles, dash-dotted line), 2.125 {\AA} (filled triangles,
thick long-dashed line), 1.9 {\AA} (filled squares, thin continuous line),
1.75 {\AA} (filled
circles, thin dashed line), concentration-dependent $\delta$ according
to Onsager--Samaras theory \protect\cite{onsager,bertrandth}
(open diamonds, thin long-dashed line).
The thick continuous line marks the critical wetting transition temperatures
as calculated from the Hamaker constant.}
\end{figure}

\end{document}